\documentclass[12pt]{article}
\usepackage{amsmath,amssymb,amsthm,amsxtra,overpic,bbm,bm,epsfig,ulem}
\usepackage{color}
\usepackage{cite}

\usepackage{graphicx}

\usepackage{hyperref}
\usepackage{url}
\usepackage{multirow}
\usepackage{textcomp}

\def\thefootnote{\fnsymbol{footnote}}

\begin{document}

\vspace{0.2cm}

\begin{center}
{\large\bf Potential divergence in tracing $\mu$ and $\tau$ flavors of astrophysical neutrinos}
\end{center}

\vspace{0.2cm}

\begin{center}
{\bf Zhi-zhong Xing$^{1,2,3}$}
\footnote{E-mail: xingzz@ihep.ac.cn}
\\
{\small
$^{1}$Institute of High Energy Physics, Chinese Academy of Sciences, Beijing 100049, China \\
$^{2}$School of Physical Sciences,
University of Chinese Academy of Sciences, Beijing 100049, China \\
$^{3}$Center of High Energy Physics, Peking University, Beijing 100871, China}
\end{center}

\begin{abstract}
We derive general formulas for three flavor fractions
$(\eta^{}_e , \eta^{}_\mu , \eta^{}_\tau)$ of the
high-energy neutrinos originating from a remote astrophysical source by using
their flavor ratios $(f^{}_e , f^{}_\mu , f^{}_\tau)$ observed at a neutrino
telescope, and diagnose a potential divergence associated with $\eta^{}_\mu$
and $\eta^{}_\tau$ as an unavoidable consequence of the $\mu$-$\tau$
interchange symmetry exhibiting in the $3\times 3$ lepton flavor mixing matrix
$U$. We present a complete set of analytical expressions for
$(\eta^{}_e , \eta^{}_\mu , \eta^{}_\tau)$ as functions of two typical $\mu$-$\tau$
symmetry breaking parameters in the standard parametrization of $U$, and
apply it to the recent IceCube all-sky neutrino flux data ranging from 5 TeV
to 10 PeV in the assumption that the relevant sources have a common flavor
composition. We also explain why only $\eta^{}_e$ and $\eta^{}_\mu + \eta^{}_\tau$
can be extracted from a precision measurement of $f^{}_e$ and $f^{}_\mu = f^{}_\tau$
in the exact $\mu$-$\tau$ flavor symmetry limit.
\end{abstract}

\newpage

\def\thefootnote{\arabic{footnote}}
\setcounter{footnote}{0}
\setcounter{figure}{0}

\section{Introduction}

It is known that the flavor oscillations of high-energy astrophysical neutrinos
over a sufficiently long distance will lose their energy- and baseline-dependent
information, but may allow a neutrino telescope to trace their flavor composition
at the source~\cite{Learned:2000sw,Halzen:2002pg,Beacom:2003nh,Becker:2007sv}.
The key point is that there exists a linear relation between the original neutrino
fluxes of different flavors $(\phi^{}_e , \phi^{}_\mu , \phi^{}_\tau)$
and the observed ones $(\Phi^{}_e , \Phi^{}_\mu , \Phi^{}_\tau)$ in the standard
three-flavor oscillation scheme
\footnote{Note that $\phi^{}_e$ (or $\Phi^{}_e$) stands for a sum of both the electron
neutrino flux and the electron antineutrino flux at the astrophysical source (or the
telescope), so do $\phi^{}_\mu$ and $\phi^{}_\tau$ (or $\Phi^{}_\mu$ and $\Phi^{}_\tau$).}:
\begin{eqnarray}
\left[\begin{matrix}
\Phi^{}_e \cr \Phi^{}_\mu \cr \Phi^{}_\tau \end{matrix}\right]
= \left(\begin{matrix}
P^{}_{e e} & P^{}_{e \mu} & P^{}_{e \tau} \cr
P^{}_{\mu e} & P^{}_{\mu \mu} & P^{}_{\mu \tau} \cr
P^{}_{\tau e} & P^{}_{\tau \mu} & P^{}_{\tau \tau}
\end{matrix}\right)
\left[\begin{matrix} \phi^{}_e \cr \phi^{}_\mu \cr
\phi^{}_\tau \end{matrix}\right] \; ,
\label{1}
\end{eqnarray}
in which the oscillation probabilities are simple functions of the elements
of the Pontecorvo-Maki-Nakagawa-Sakata (PMNS) lepton flavor mixing matrix
$U$~\cite{Pontecorvo:1957cp,Maki:1962mu,Pontecorvo:1967fh} and satisfy
\begin{eqnarray}
P^{}_{\alpha\beta} = P^{}_{\beta\alpha} = \sum^3_{i=1}
\left(|U^{}_{\alpha i}|^2 \hspace{0.05cm} |U^{}_{\beta i}|^2\right) \; ,
\label{2}
\end{eqnarray}
where the Greek subscripts $\alpha$ and $\beta$ run over the flavor indices
$e$, $\mu$ and $\tau$ for three active neutrinos $\nu^{}_e$, $\nu^{}_\mu$
and $\nu^{}_\tau$. The unitarity of $U$ assures
$\Phi^{}_0 \equiv \Phi^{}_e + \Phi^{}_\mu + \Phi^{}_\tau
= \phi^{}_e + \phi^{}_\mu + \phi^{}_\tau$ to hold, implying
that the total neutrino flux $\Phi^{}_0$ is conserved. In practice, the
measurements of $\Phi^{}_\alpha$ (for $\alpha = e, \mu, \tau$) and
$\Phi^{}_0$ are expected to involve quite large systematic uncertainties,
but such uncertainties can be largely cancelled out in the flux ratios
\begin{eqnarray}
f^{}_e \equiv \frac{\Phi^{}_e}{\Phi^{}_0} \; , \;\;
f^{}_\mu \equiv \frac{\Phi^{}_\mu}{\Phi^{}_0} \; , \;\;
f^{}_\tau \equiv \frac{\Phi^{}_\tau}{\Phi^{}_0} \; ,
\label{3}
\end{eqnarray}
which characterize the three-flavor composition of
astrophysical neutrinos observed at a neutrino telescope like the
running IceCube~\cite{Abbasi:2025fjc} and KM3NeT~\cite{KM3NeT:2025npi}
detectors. A precision measurement of these three flavor ratios will help
determine the original neutrino flavor fractions
\begin{eqnarray}
\eta^{}_e \equiv \frac{\phi^{}_e}{\Phi^{}_0} \; , \;\;
\eta^{}_\mu \equiv \frac{\phi^{}_\mu}{\Phi^{}_0} \; , \;\;
\eta^{}_\tau \equiv \frac{\phi^{}_\tau}{\Phi^{}_0} \; ,
\label{4}
\end{eqnarray}
for a remote astrophysical source. For example, $\eta^{}_e = 1/3$,
$\eta^{}_\mu = 2/3$ and $\eta^{}_\tau = 0$ have been conjectured for
those astrophysical neutrinos produced from the decays of charged pions
that originate from very energetic proton-proton and proton-photon
collisions~\cite{Learned:2000sw,Halzen:2002pg,Beacom:2003nh,Becker:2007sv},
but whether such a conjecture is true or not needs an experimental test in
the precision era of neutrino astronomy.

This paper is intended to derive the general and explicit
expressions of $\eta^{}_\alpha$ in terms of $f^{}_\beta$,
$|U^{}_{\alpha i}|^2$ and $|U^{}_{\beta i}|^2$ (for $i = 1, 2, 3$ and
$\alpha, \beta = e, \mu, \tau$), so as to provide an analytical tool to
infer the possible flavor composition of an astrophysical source from a
precision measurement of the corresponding flavor distribution at a
neutrino telescope. Such a complete calculation has not been done, although
a preliminary analysis along this line of thought was carried out by
Fu {\it et al} in Refs.~\cite{Fu:2012zr,Fu:2014isa}
\footnote{They pointed out a potential divergence of $\eta^{}_\alpha$ inferred
from $f^{}_\beta$ (for $\alpha, \beta = e, \mu, \tau$) if the equalities
$|U^{}_{\mu i}| = |U^{}_{\tau i}|$ hold in the $\mu$-$\tau$ interchange
symmetry limit (for $i = 1, 2, 3$), and illustrated this interesting
observation by typically taking the ``tribimaximal" flavor mixing
pattern of $U$~\cite{Harrison:2002er,Xing:2002sw,He:2003rm} for example.}.
The present work is remarkably different from theirs, at least in the following
aspects.
\begin{itemize}
\item     We work out the concrete analytical expressions of $\eta^{}_\alpha$
as functions of $f^{}_\beta$, $|U^{}_{\alpha i}|^2$ and $|U^{}_{\beta i}|^2$
without making any special assumptions. These results clearly indicate that
$\eta^{}_\mu$ and $\eta^{}_\tau$ may diverge to infinity in the exact
$\mu$-$\tau$ symmetry case (i.e., $|U^{}_{\mu i}| = |U^{}_{\tau i}|$ for
$i = 1, 2, 3$), but $\eta^{}_e$ is insensitive to this striking
flavor symmetry to a large extent.

\item     To be more explicit, we diagnose where and to what extent the
divergence of $\eta^{}_\mu$ and $\eta^{}_\tau$ may emerge by expressing
them in terms of two {\it small} $\mu$-$\tau$ symmetry breaking parameters
\footnote{Given the facts that $\theta^{}_{23}$ is very close to $\pi/4$ and
$\theta^{}_{13}$ is of ${\cal O}(0.1)$~\cite{ParticleDataGroup:2024cfk},
both $\epsilon^{}_\theta$ and $\epsilon^{}_\delta$ are expected to be small
enough no matter whether the CP-violating phase $\delta^{}_\nu$ is finally
verified to be close to $\pm\pi/2$ or not.}
\begin{eqnarray}
\epsilon^{}_\theta \equiv \cos 2\theta^{}_{23} \; , \quad
\epsilon^{}_\delta \equiv \sin\theta^{}_{13} \sin 2\theta^{}_{23}
\cos\delta^{}_\nu \; \hspace{1cm}
\label{5}
\end{eqnarray}
in the standard parametrization of the PMNS matrix
$U$~\cite{ParticleDataGroup:2024cfk}. The result of $\eta^{}_e$
is also obtained, and its weaker sensitivity to $\epsilon^{}_\theta$
or $\epsilon^{}_\delta$ is discussed as well. This set of new formulas will
be very useful to experimentally test the intrinsic correlation between
$(\eta^{}_e , \eta^{}_\mu , \eta^{}_\tau)$ and $(f^{}_e , f^{}_\mu , f^{}_\tau)$
by combining the neutrino telescope data with the neutrino oscillation data.

\item    We apply our formulas to the recently reported IceCube all-sky
neutrino flux data ranging from 5 TeV to 10 PeV~\cite{Abbasi:2025fjc}
by assuming that the relevant astrophysical sources have the same flavor
composition. In this case we find that the IceCube best-fit values of
$f^{}_e$, $f^{}_\mu$ and $f^{}_\tau$ lead to reasonable results of
$\eta^{}_e$ and $\eta^{}_\mu + \eta^{}_\tau$, but the individual values
of $\eta^{}_\mu$ and $\eta^{}_\tau$ do not make sense because they suffer
from the divergent effects caused by both an approximate $\mu$-$\tau$
flavor symmetry of $U$ and the experimental uncertainties of $f^{}_\mu$
and $f^{}_\tau$.
\end{itemize}
In addition, we explain why only $\eta^{}_e$ and $\eta^{}_\mu + \eta^{}_\tau$
at an astrophysical source can be extracted from the measurements of
$f^{}_e$ and $f^{}_\mu = f^{}_\tau$ at a neutrino telescope if the
$\mu$-$\tau$ flavor symmetry is exact, and the latest JUNO and Daya Bay
precision antineutrino oscillation data on the flavor mixing angles
$\theta^{}_{12}$~\cite{JUNO:2025gmd} and $\theta^{}_{13}$~\cite{DayaBay:2022orm}
will help a lot in this extreme case.

The remaining parts of this paper is organized as follows. In section 2
we derive general formulas for the source flavor fractions
$(\eta^{}_e , \eta^{}_\mu , \eta^{}_\tau)$ in terms of the observed
flavor ratios $(f^{}_e , f^{}_\mu , f^{}_\tau)$ and the moduli of the
PMNS matrix elements. Section 3 is devoted to formulating
$(\eta^{}_e , \eta^{}_\mu , \eta^{}_\tau)$ as functions
of the $\mu$-$\tau$ symmetry breaking parameters $\epsilon^{}_\theta$
and $\epsilon^{}_\delta$, and to illustrating their sensitivities
to small $\epsilon^{}_\theta$ and $\epsilon^{}_\delta$ by use
of the recent IceCube all-sky neutrino flux data ranging from
5 TeV to 10 PeV. In section 4 we show that it is only
possible to extract $\eta^{}_e$ and $\eta^{}_\mu + \eta^{}_\tau$, rather
than individual $\eta^{}_\mu$ and $\eta^{}_\tau$, from the measurements
of $f^{}_e$ and $f^{}_\mu = f^{}_\tau$ in the exact $\mu$-$\tau$ flavor
symmetry limit. Section 5 is devoted to a brief summary of our main results.

\section{Inferring the initial flavor fractions}

Let us begin with a normalization of the neutrino fluxes appearing in
Eq.~(\ref{1}) with the total neutrino flux $\Phi^{}_0$, as done in
Eqs.~(\ref{3}) and (\ref{4}). Then we are left with a system of three
linear equations in three variables $\eta^{}_e$, $\eta^{}_\mu$ and
$\eta^{}_\tau$ as follows:
\begin{eqnarray}
\left(\begin{matrix}
P^{}_{e e} & P^{}_{e \mu} & P^{}_{e \tau} \cr
P^{}_{\mu e} & P^{}_{\mu \mu} & P^{}_{\mu \tau} \cr
P^{}_{\tau e} & P^{}_{\tau \mu} & P^{}_{\tau \tau}
\end{matrix}\right)
\left[\begin{matrix} \eta^{}_e \cr \eta^{}_\mu \cr
\eta^{}_\tau \end{matrix}\right]
=
\left[\begin{matrix} f^{}_e \cr f^{}_\mu \cr
f^{}_\tau \end{matrix}\right] \; .
\label{6}
\end{eqnarray}
Following Gabriel Cramer's rule, one may solve the above equations by
calculating the determinants
\begin{eqnarray}
D^{}_0 \hspace{-0.2cm} & = & \hspace{-0.2cm} \left| \begin{matrix}
P^{}_{ee} & P^{}_{e\mu} & P^{}_{e\tau} \cr
P^{}_{\mu e} & P^{}_{\mu\mu} & P^{}_{\mu\tau} \cr
P^{}_{\tau e} & P^{}_{\tau\mu} & P^{}_{\tau\tau} \cr
\end{matrix} \right| \; ,
\nonumber \\
D^{}_e \hspace{-0.2cm} & = & \hspace{-0.2cm} \left| \begin{matrix}
f^{}_{e} & \hspace{0.12cm} P^{}_{e\mu} \hspace{0.12cm} & P^{}_{e\tau} \cr
f^{}_{\mu} & P^{}_{\mu\mu} & P^{}_{\mu\tau} \cr
f^{}_{\tau} & P^{}_{\tau\mu} & P^{}_{\tau\tau} \cr
\end{matrix} \right| \; ,
\nonumber \\
D^{}_\mu \hspace{-0.2cm} & = & \hspace{-0.2cm} \left| \begin{matrix}
P^{}_{ee} & \hspace{0.14cm} f^{}_{e} \hspace{0.14cm} & P^{}_{e\tau} \cr
P^{}_{\mu e} & f^{}_{\mu} & P^{}_{\mu\tau} \cr
P^{}_{\tau e} & f^{}_{\tau} & P^{}_{\tau\tau} \cr
\end{matrix} \right| \; ,
\nonumber \\
D^{}_\tau \hspace{-0.2cm} & = & \hspace{-0.2cm} \left| \begin{matrix}
P^{}_{ee} & \hspace{0.127cm} P^{}_{e\mu} \hspace{0.127cm} & f^{}_{e} \cr
P^{}_{\mu e} & P^{}_{\mu\mu} & f^{}_{\mu} \cr
P^{}_{\tau e} & P^{}_{\tau\mu} & f^{}_{\tau} \cr
\end{matrix} \right| \; . \hspace{0.7cm}
\label{7}
\end{eqnarray}
Under the condition of $D^{}_0 \neq 0$, the solutions to Eq.~(\ref{6})
turn out to be
\begin{eqnarray}
\eta^{}_e = \frac{D^{}_e}{D^{}_0} \; , \;\;
\eta^{}_\mu = \frac{D^{}_\mu}{D^{}_0} \; , \;\;
\eta^{}_\tau = \frac{D^{}_\tau}{D^{}_0} \; .
\label{8}
\end{eqnarray}
It is obvious that $f^{}_e = f^{}_\mu = f^{}_\tau = 1/3$ corresponds
to a trivial solution $\eta^{}_e = \eta^{}_\mu = \eta^{}_\tau =1/3$ as
guaranteed by the unitarity of $U$ or the conservation of probability for
neutrino oscillations. This solution is by no means interesting as it is
unrealistic and fully independent of the PMNS flavor mixing parameters.
In contrast,
an exact $\mu$-$\tau$ interchange symmetry associated with the second and
third rows of the PMNS matrix $U$ leads to a well-known nontrivial result
$f^{}_e = f^{}_\mu = f^{}_\tau = 1/3$ at a neutrino telescope provided
the astrophysical source has $\eta^{}_e = 1/3$, $\eta^{}_\mu = 2/3$ and
$\eta^{}_\tau = 0$ for its
flavor composition~\cite{Xing:2006xd,Xing:2006uk,Xing:2008fg,Xing:2011zm}
(some more comprehensive analyses can be found, e.g., in  Refs.~\cite{Bustamante:2015waa,Bustamante:2019sdb,Song:2020nfh,
Testagrossa:2023ukh,Coleman:2024scd}).
But here we focus on the general solutions to Eq.~(\ref{6}).

Taking account of Eq.~(\ref{2}), we get the explicit expression of $D^{}_0$
after a straightforward calculation:
\begin{eqnarray}
D^{}_0 \hspace{-0.2cm} & = & \hspace{-0.2cm}
\Big[ |U^{}_{e 1}|^2 \left(|U^{}_{\mu 2}|^2
|U^{}_{\tau 3}|^2 - |U^{}_{\mu 3}|^2
|U^{}_{\tau 2}|^2\right)
+ |U^{}_{e 2}|^2 \left(|U^{}_{\mu 3}|^2
|U^{}_{\tau 1}|^2 - |U^{}_{\mu 1}|^2
|U^{}_{\tau 3}|^2\right)
\nonumber \\
\hspace{-0.2cm} & & \hspace{-0.2cm} +
\hspace{0.1cm} |U^{}_{e 3}|^2 \left(|U^{}_{\mu 1}|^2
|U^{}_{\tau 2}|^2 - |U^{}_{\mu 2}|^2
|U^{}_{\tau 1}|^2\right) \Big]^2 \; .
\label{9}
\end{eqnarray}
We see that $D^{}_0 \geq 0$ must hold. Whether $D^{}_0 = 0$ could result
from an exact cancellation among the three terms
of $D^{}_0$ remains an open question, as the moduli of the nine PMNS matrix
elements have not been determined to a sufficiently good degree of
accuracy~\cite{ParticleDataGroup:2024cfk,Capozzi:2025wyn,Capozzi:2025ovi,
Esteban:2024eli,Esteban:2026phq}.
A more interesting possibility for $D^{}_0 = 0$ arises from the exact
$\mu$-$\tau$ interchange flavor symmetry~\cite{Xing:2015fdg,Xing:2022uax}
\begin{eqnarray}
|U^{}_{\mu 1}| = |U^{}_{\tau 1}| \; , ~~
|U^{}_{\mu 2}| = |U^{}_{\tau 2}| \; , ~~
|U^{}_{\mu 3}| = |U^{}_{\tau 3}| \; ,
\label{10}
\end{eqnarray}
which is a natural consequence of a lot of neutrino mass models
based on simple flavor symmetry groups~\cite{Xing:2019vks}. This possibility
will be discussed in some detail later on.

Working out the concrete expressions of $D^{}_\alpha$ (for $\alpha = e, \mu,
\tau$) in Eq.~(\ref{7}), we arrive at the analytical results of $(\eta^{}_e,
\eta^{}_\mu, \eta^{}_\tau)$ in terms of $(f^{}_e , f^{}_\mu , f^{}_\tau)$
and the PMNS moduli as follows:
\begin{eqnarray}
\left[\begin{matrix} \eta^{}_e \cr \eta^{}_\mu \cr
\eta^{}_\tau \end{matrix}\right]
= \frac{1}{D^{}_0}
\left(\begin{matrix}
C^{}_{e e} & C^{}_{e \mu} & C^{}_{e \tau} \cr
C^{}_{e \mu} & C^{}_{\mu \mu} & C^{}_{\mu \tau} \cr
C^{}_{e \tau} & C^{}_{\mu \tau} & C^{}_{\tau \tau}
\end{matrix}\right)
\left[\begin{matrix} f^{}_e \cr f^{}_\mu \cr
f^{}_\tau \end{matrix}\right] \; ,
\label{11}
\end{eqnarray}
where
\begin{eqnarray}
C^{}_{ee} \hspace{-0.2cm} & = & \hspace{-0.2cm}
\left(|U^{}_{\mu 1}|^2 |U^{}_{\tau 2}|^2 -
|U^{}_{\mu 2}|^2 |U^{}_{\tau 1}|^2\right)^2
+ \left(|U^{}_{\mu 2}|^2 |U^{}_{\tau 3}|^2 -
|U^{}_{\mu 3}|^2 |U^{}_{\tau 2}|^2\right)^2
\nonumber \\
\hspace{-0.2cm} & & \hspace{-0.2cm}
+ \left(|U^{}_{\mu 3}|^2 |U^{}_{\tau 1}|^2 -
|U^{}_{\mu 1}|^2 |U^{}_{\tau 3}|^2\right)^2 \; ,
\nonumber \\
C^{}_{\mu\mu} \hspace{-0.2cm} & = & \hspace{-0.2cm}
\left(|U^{}_{\tau 1}|^2 |U^{}_{e 2}|^2 -
|U^{}_{\tau 2}|^2 |U^{}_{e 1}|^2\right)^2
+ \left(|U^{}_{\tau 2}|^2 |U^{}_{e 3}|^2 -
|U^{}_{\tau 3}|^2 |U^{}_{e 2}|^2\right)^2
\nonumber \\
\hspace{-0.2cm} & & \hspace{-0.2cm}
+ \left(|U^{}_{\tau 3}|^2 |U^{}_{e 1}|^2 -
|U^{}_{\tau 1}|^2 |U^{}_{e 3}|^2\right)^2 \; ,
\nonumber \\
C^{}_{\tau\tau} \hspace{-0.2cm} & = & \hspace{-0.2cm}
\left(|U^{}_{e 1}|^2 |U^{}_{\mu 2}|^2 -
|U^{}_{e 2}|^2 |U^{}_{\mu 1}|^2\right)^2
+ \left(|U^{}_{e 2}|^2 |U^{}_{\mu 3}|^2 -
|U^{}_{e 3}|^2 |U^{}_{\mu 2}|^2\right)^2 \hspace{0.5cm}
\nonumber \\
\hspace{-0.2cm} & & \hspace{-0.2cm}
+ \left(|U^{}_{e 3}|^2 |U^{}_{\mu 1}|^2 -
|U^{}_{e 1}|^2 |U^{}_{\mu 3}|^2\right)^2 \; ,
\label{12}
\end{eqnarray}
together with
\begin{eqnarray}
C^{}_{e\mu} \hspace{-0.2cm} & = & \hspace{-0.2cm}
\left(|U^{}_{\mu 1}|^2 |U^{}_{\tau 2}|^2 -
|U^{}_{\mu 2}|^2 |U^{}_{\tau 1}|^2\right)
\left(|U^{}_{\tau 1}|^2 |U^{}_{e 2}|^2 -
|U^{}_{\tau 2}|^2 |U^{}_{e 1}|^2\right)
\nonumber \\
\hspace{-0.2cm} & & \hspace{-0.2cm}
+ \left(|U^{}_{\mu 2}|^2 |U^{}_{\tau 3}|^2 -
|U^{}_{\mu 3}|^2 |U^{}_{\tau 2}|^2\right)
\left(|U^{}_{\tau 2}|^2 |U^{}_{e 3}|^2 -
|U^{}_{\tau 3}|^2 |U^{}_{e 2}|^2\right)
\nonumber \\
\hspace{-0.2cm} & & \hspace{-0.2cm}
+ \left(|U^{}_{\mu 3}|^2 |U^{}_{\tau 1}|^2 -
|U^{}_{\mu 1}|^2 |U^{}_{\tau 3}|^2\right)
\left(|U^{}_{\tau 3}|^2 |U^{}_{e 1}|^2 -
|U^{}_{\tau 1}|^2 |U^{}_{e 3}|^2\right) \; ,
\nonumber \\
C^{}_{e\tau} \hspace{-0.2cm} & = & \hspace{-0.2cm}
\left(|U^{}_{e 1}|^2 |U^{}_{\mu 2}|^2 -
|U^{}_{e 2}|^2 |U^{}_{\mu 1}|^2\right)
\left(|U^{}_{\mu 1}|^2 |U^{}_{\tau 2}|^2 -
|U^{}_{\mu 2}|^2 |U^{}_{\tau 1}|^2\right)
\nonumber \\
\hspace{-0.2cm} & & \hspace{-0.2cm}
+ \left(|U^{}_{e 2}|^2 |U^{}_{\mu 3}|^2 -
|U^{}_{e 3}|^2 |U^{}_{\mu 2}|^2\right)
\left(|U^{}_{\mu 2}|^2 |U^{}_{\tau 3}|^2 -
|U^{}_{\mu 3}|^2 |U^{}_{\tau 2}|^2\right)
\nonumber \\
\hspace{-0.2cm} & & \hspace{-0.2cm}
+ \left(|U^{}_{e 3}|^2 |U^{}_{\mu 1}|^2 -
|U^{}_{e 1}|^2 |U^{}_{\mu 3}|^2\right)
\left(|U^{}_{\mu 3}|^2 |U^{}_{\tau 1}|^2 -
|U^{}_{\mu 1}|^2 |U^{}_{\tau 3}|^2\right) \; ,
\nonumber \\
C^{}_{\mu\tau} \hspace{-0.2cm} & = & \hspace{-0.2cm}
\left(|U^{}_{\tau 1}|^2 |U^{}_{e 2}|^2 -
|U^{}_{\tau 2}|^2 |U^{}_{e 1}|^2\right)
\left(|U^{}_{e 1}|^2 |U^{}_{\mu 2}|^2 -
|U^{}_{e 2}|^2 |U^{}_{\mu 1}|^2\right)
\nonumber \\
\hspace{-0.2cm} & & \hspace{-0.2cm}
+ \left(|U^{}_{\tau 2}|^2 |U^{}_{e 3}|^2 -
|U^{}_{\tau 3}|^2 |U^{}_{e 2}|^2\right)
\left(|U^{}_{e 2}|^2 |U^{}_{\mu 3}|^2 -
|U^{}_{e 3}|^2 |U^{}_{\mu 2}|^2\right)
\nonumber \\
\hspace{-0.2cm} & & \hspace{-0.2cm}
+ \left(|U^{}_{\tau 3}|^2 |U^{}_{e 1}|^2 -
|U^{}_{\tau 1}|^2 |U^{}_{e 3}|^2\right)
\left(|U^{}_{e 3}|^2 |U^{}_{\mu 1}|^2 -
|U^{}_{e 1}|^2 |U^{}_{\mu 3}|^2\right) \; . \hspace{0.55cm}
\label{13}
\end{eqnarray}
Note that these results are new and instructive, and they are especially
useful for us to understand why the source flavor fractions
$\eta^{}_\alpha$ (for $\alpha = e, \mu, \tau$) are sensitive (or insensitive)
to the $\mu$-$\tau$ interchange symmetry of $U$. Three immediate comments are
in order.
\begin{itemize}
\item     Given the exact $\mu$-$\tau$ interchange flavor symmetry in Eq.~(\ref{10}),
one will be simply left with $C^{}_{ee} = C^{}_{e\mu} = C^{}_{e\tau} = 0$
in addition to the relation $C^{}_{\mu\mu} = C^{}_{\tau\tau} = -C^{}_{\mu\tau}$.

\item     It is obvious that $C^{}_{ee}$ {\it quadratically} approaches zero as $D^{}_0$
does if the $\mu$-$\tau$ symmetry conditions in Eq.~(\ref{10}) are satisfied. In this
case the ratio $C^{}_{ee}/D^{}_0$ should be finite unless $D^{}_0$ itself vanishes,
as can be more clearly
seen in section 3 after $C^{}_{\alpha\beta}$ and $D^{}_0$ are explicitly expressed
in terms of the Euler-like flavor mixing angles of $U$ and its $\mu$-$\tau$ symmetry
breaking parameters $\epsilon^{}_\theta$ and $\epsilon^{}_\delta$. Namely, both
$C^{}_{ee}$ and $D^{}_0$ are linear combinations of the terms proportional to
$\epsilon^2_\theta$, $\epsilon^2_\delta$ and $\epsilon^{}_\theta \epsilon^{}_\delta$.

\item     In comparison, Eqs.~(\ref{12}) and (\ref{13}) tell us that $C^{}_{e\mu}$ and
$C^{}_{e\tau}$ {\it linearly} approach zero in the $\mu$-$\tau$ flavor symmetry limit
(i.e., their leading terms are proportional to $\epsilon^{}_\theta$ and $\epsilon^{}_\delta$),
while $C^{}_{\mu\mu}$, $C^{}_{\mu\tau}$ and $C^{}_{\tau\tau}$ are in general finite
in the same limit (i.e., their leading terms are independent of $\epsilon^{}_\theta$ and
$\epsilon^{}_\delta$). That is why the potential divergence of $C^{}_{\mu\mu}/D^{}_0$,
$C^{}_{\mu\tau}/D^{}_0$ and $C^{}_{\tau\tau}/D^{}_0$ is expected to be much stronger
than that of $C^{}_{e\mu}/D^{}_0$ and $C^{}_{e\tau}/D^{}_0$ for vanishing or vanishingly
small $\epsilon^{}_\theta$ and $\epsilon^{}_\delta$, leading consequently to an
unavoidable divergence of $\eta^{}_\mu$ and $\eta^{}_\tau$.
\end{itemize}
To diagnose where and to what extent the potential divergence of $\eta^{}_\alpha$ (for
$\alpha = e, \mu, \tau$) may emerge
in a more explicit way, let us proceed to reexpress Eqs.~(\ref{9}), (\ref{12}) and
(\ref{13}) by means of $\epsilon^{}_\theta$ and $\epsilon^{}_\delta$ in the standard
parametrization of $U$.

\section{Diagnosing the origin of divergence}

As the nine moduli of the PMNS matrix elements are independent of possible Majorana
CP-violating phases, they can be expressed in terms of the three flavor mixing angles
$(\theta^{}_{12}, \theta^{}_{13}, \theta^{}_{23}$) and the Dirac CP-violating phase
($\delta^{}_\nu$) in the standard parametrization of $U$~\cite{ParticleDataGroup:2024cfk}
as follows:
\begin{eqnarray}
|U^{}_{e 1}|^2 \hspace{-0.2cm} & = & \hspace{-0.2cm}
c^2_{12} c^2_{13} \; , \quad
|U^{}_{e 2}|^2 = s^2_{12} c^2_{13} \; , \quad
|U^{}_{e 3}|^2 = s^2_{13} \; ,
\nonumber \\
|U^{}_{\mu 1}|^2 \hspace{-0.2cm} & = & \hspace{-0.2cm}
\frac{1}{2} s^2_{12} \left(1 + \epsilon^{}_\theta\right)
+ \frac{1}{2} c^2_{12} s^2_{13} \left(1 - \epsilon^{}_\theta\right)
+ c^{}_{12} s^{}_{12} \epsilon^{}_\delta \; ,
\nonumber \\
|U^{}_{\mu 2}|^2 \hspace{-0.2cm} & = & \hspace{-0.2cm}
\frac{1}{2} c^2_{12} \left(1 + \epsilon^{}_\theta\right)
+ \frac{1}{2} s^2_{12} s^2_{13} \left(1 - \epsilon^{}_\theta\right)
- c^{}_{12} s^{}_{12} \epsilon^{}_\delta \; ,
\nonumber \\
|U^{}_{\tau 1}|^2 \hspace{-0.2cm} & = & \hspace{-0.2cm}
\frac{1}{2} s^2_{12} \left(1 - \epsilon^{}_\theta\right)
+ \frac{1}{2} c^2_{12} s^2_{13} \left(1 + \epsilon^{}_\theta\right)
- c^{}_{12} s^{}_{12} \epsilon^{}_\delta \; ,
\nonumber \\
|U^{}_{\tau 2}|^2 \hspace{-0.2cm} & = & \hspace{-0.2cm}
\frac{1}{2} c^2_{12} \left(1 - \epsilon^{}_\theta\right)
+ \frac{1}{2} s^2_{12} s^2_{13} \left(1 + \epsilon^{}_\theta\right)
+ c^{}_{12} s^{}_{12} \epsilon^{}_\delta \; , \hspace{0.6cm}
\nonumber \\
|U^{}_{\mu 3}|^2 \hspace{-0.2cm} & = & \hspace{-0.2cm}
\frac{1}{2} c^2_{13} \left(1 - \epsilon^{}_\theta\right) \; , \quad
|U^{}_{\tau 3}|^2 = \frac{1}{2} c^2_{13} \left(1 + \epsilon^{}_\theta\right) \; ,
\label{14}
\end{eqnarray}
where $c^{}_{ij} \equiv \cos\theta^{}_{ij}$ and $s^{}_{ij} \equiv
\sin\theta^{}_{ij}$ with $\theta^{}_{ij}$ lying in the first quadrant
(for $ij = 12, 13, 23$), and the $\mu$-$\tau$ symmetry breaking parameters
$\epsilon^{}_\theta$ and $\epsilon^{}_\delta$ have been defined in
Eq.~(\ref{5}). The necessary and sufficient conditions for an exact
$\mu$-$\tau$ reflection symmetry of $U$ are $\theta^{}_{23} = \pi/4$
and $\delta^{}_\nu = \pm \pi/2$~\cite{Xing:2015fdg,Xing:2022uax}, which
are equivalent to $\epsilon^{}_\theta = \epsilon^{}_\delta = 0$.
Although the octant of $\theta^{}_{23}$ and the quadrant of $\delta^{}_\nu$
have not been fixed by current neutrino oscillation
experiments~\cite{ParticleDataGroup:2024cfk,Capozzi:2025wyn,Capozzi:2025ovi,
Esteban:2024eli,Esteban:2026phq},
the smallness of $\epsilon^{}_\theta$ and $\epsilon^{}_\delta$ in magnitude
is definitely guaranteed. In other words, the observed pattern of lepton
flavor mixing at least exhibits an approximate $\mu$-$\tau$ flavor symmetry.

Substituting Eq.~(\ref{14}) into Eqs.~(\ref{9}), (\ref{12}) and (\ref{13}),
we obtain the following expressions of $D^{}_0$ and $C^{}_{\alpha\beta}$ (for
$\alpha, \beta = e, \mu, \tau$) in terms of $\theta^{}_{12}$, $\theta^{}_{13}$,
$\epsilon^{}_\theta$ and $\epsilon^{}_\delta$:
\begin{eqnarray}
D^{}_0 = \cos^2 2\theta^{}_{12} \cos^2 2\theta^{}_{13} \epsilon^2_\theta +
\frac{1}{4} \sin^2 2\theta^{}_{12} \left(1 - 3 s^2_{13}\right)^2
\epsilon^2_\delta
- \frac{1}{2} \sin 4\theta^{}_{12} \cos 2\theta^{}_{13}
\left(1 - 3 s^2_{13}\right) \epsilon^{}_\theta \epsilon^{}_\delta \; ,
\label{15}
\end{eqnarray}
and
\begin{eqnarray}
C^{}_{ee} \hspace{-0.2cm} & = & \hspace{-0.2cm}
\left[\cos^2 2\theta^{}_{12} \left(1 - 2 s^2_{13} + 2 s^4_{13}\right)
+ \frac{1}{2} \sin^2 2\theta^{}_{12} c^4_{13}\right] \epsilon^2_\theta +
\frac{1}{4} \sin^2 2\theta^{}_{12} \left(3 - 2 s^2_{13} + 3 s^4_{13}\right)
\epsilon^2_\delta \hspace{0.95cm}
\nonumber \\
\hspace{-0.2cm} & & \hspace{-0.2cm}
- \hspace{0.1cm} \frac{1}{2} \sin 4\theta^{}_{12} \left(1 - s^2_{13}
+ 2 s^4_{13}\right) \epsilon^{}_\theta \epsilon^{}_\delta \; ,
\nonumber \\
C^{}_{e\mu} \hspace{-0.2cm} & = & \hspace{-0.2cm}
\frac{1}{4} \left[\cos^2 2\theta^{}_{12} \sin^2 2\theta^{}_{13}
- \sin^2 2\theta^{}_{12} \left(1 - 4 s^2_{13} + 3 s^4_{13}\right)\right]
\epsilon^{}_\theta
- \frac{1}{4} \sin 4\theta^{}_{12} c^2_{13} \epsilon^{}_\delta
\nonumber \\
\hspace{-0.2cm} & & \hspace{-0.2cm}
- \hspace{0.1cm} \frac{1}{4} \left( \cos^2 2\theta^{}_{12} \sin^2 2\theta^{}_{13}
+ \sin^2 2\theta^{}_{12} c^4_{13}\right) \epsilon^2_\theta
- \frac{1}{4} \sin^2 2\theta^{}_{12} \left(1 + 2 s^2_{13} - 3 s^4_{13}\right)
\epsilon^2_\delta
\nonumber \\
\hspace{-0.2cm} & & \hspace{-0.2cm}
+ \hspace{0.1cm} \frac{1}{4} \sin 4\theta^{}_{12} \sin^2 2\theta^{}_{13}
\epsilon^{}_\theta \epsilon^{}_\delta \; ,
\nonumber \\
C^{}_{e\tau} \hspace{-0.2cm} & = & \hspace{-0.2cm}
\frac{1}{4} \left[-\cos^2 2\theta^{}_{12} \sin^2 2\theta^{}_{13}
+ \sin^2 2\theta^{}_{12} \left(1 - 4 s^2_{13} + 3 s^4_{13}\right)\right]
\epsilon^{}_\theta
+ \frac{1}{4} \sin 4\theta^{}_{12} c^2_{13} \epsilon^{}_\delta
\nonumber \\
\hspace{-0.2cm} & & \hspace{-0.2cm}
- \hspace{0.1cm} \frac{1}{4} \left( \cos^2 2\theta^{}_{12} \sin^2 2\theta^{}_{13}
+ \sin^2 2\theta^{}_{12} c^4_{13}\right) \epsilon^2_\theta
- \frac{1}{4} \sin^2 2\theta^{}_{12} \left(1 + 2 s^2_{13} - 3 s^4_{13}\right)
\epsilon^2_\delta
\nonumber \\
\hspace{-0.2cm} & & \hspace{-0.2cm}
+ \hspace{0.1cm} \frac{1}{4} \sin 4\theta^{}_{12} \sin^2 2\theta^{}_{13}
\epsilon^{}_\theta \epsilon^{}_\delta \; ,
\nonumber \\
C^{}_{\mu\mu} \hspace{-0.2cm} & = & \hspace{-0.2cm}
\frac{1}{2} \cos^2 2\theta^{}_{12} \left(1 - 3 s^2_{13} + 3 s^4_{13}\right)
+ \frac{1}{8} \sin^2 2\theta^{}_{12} \left(1 - 6 s^2_{13} + 9 s^4_{13}\right)
\nonumber \\
\hspace{-0.2cm} & & \hspace{-0.2cm}
- \hspace{0.1cm} \frac{1}{4} \left[\cos^2 2\theta^{}_{12} \sin^2 2\theta^{}_{13}
+ \sin^2 2\theta^{}_{12} \left(1 - 4 s^2_{13} + 3 s^4_{13}\right)\right]
\epsilon^{}_\theta
+ \frac{1}{4} \sin 4\theta^{}_{12} c^2_{13} \epsilon^{}_\delta
\nonumber \\
\hspace{-0.2cm} & & \hspace{-0.2cm}
+ \hspace{0.1cm} \frac{1}{8} \left[4 \cos^2 2\theta^{}_{12}
\left(1 - 3 s^2_{13} + 3 s^4_{13}\right) + \sin^2 2\theta^{}_{12} c^4_{13}\right]
\epsilon^2_\theta
+ \frac{1}{4} \sin^2 2\theta^{}_{12} \left(1 - 2 s^2_{13}
+ 3 s^4_{13}\right) \epsilon^2_\delta \hspace{0.1cm}
\nonumber \\
\hspace{-0.2cm} & & \hspace{-0.2cm}
- \hspace{0.1cm} \frac{1}{4} \sin 4\theta^{}_{12} \left(1 - 3 s^2_{13} + 4 s^4_{13}\right)
\epsilon^{}_\theta \epsilon^{}_\delta \; ,
\nonumber \\
C^{}_{\tau\tau} \hspace{-0.2cm} & = & \hspace{-0.2cm}
\frac{1}{2} \cos^2 2\theta^{}_{12} \left(1 - 3 s^2_{13} + 3 s^4_{13}\right)
+ \frac{1}{8} \sin^2 2\theta^{}_{12} \left(1 - 6 s^2_{13} + 9 s^4_{13}\right)
\nonumber \\
\hspace{-0.2cm} & & \hspace{-0.2cm}
+ \hspace{0.1cm} \frac{1}{4} \left[\cos^2 2\theta^{}_{12} \sin^2 2\theta^{}_{13}
+ \sin^2 2\theta^{}_{12} \left(1 - 4 s^2_{13} + 3 s^4_{13}\right)\right]
\epsilon^{}_\theta
- \frac{1}{4} \sin 4\theta^{}_{12} c^2_{13} \epsilon^{}_\delta
\nonumber \\
\hspace{-0.2cm} & & \hspace{-0.2cm}
+ \hspace{0.1cm} \frac{1}{8} \left[4 \cos^2 2\theta^{}_{12}
\left(1 - 3 s^2_{13} + 3 s^4_{13}\right) + \sin^2 2\theta^{}_{12} c^4_{13}\right]
\epsilon^2_\theta
+ \frac{1}{4} \sin^2 2\theta^{}_{12} \left(1 - 2 s^2_{13}
+ 3 s^4_{13}\right) \epsilon^2_\delta
\nonumber \\
\hspace{-0.2cm} & & \hspace{-0.2cm}
- \hspace{0.1cm} \frac{1}{4} \sin 4\theta^{}_{12} \left(1 - 3 s^2_{13} + 4 s^4_{13}\right)
\epsilon^{}_\theta \epsilon^{}_\delta \; ,
\nonumber \\
C^{}_{\mu\tau} \hspace{-0.2cm} & = & \hspace{-0.2cm}
- \frac{1}{2} \cos^2 2\theta^{}_{12} \left(1 - 3 s^2_{13} + 3 s^4_{13}\right)
- \frac{1}{8} \sin^2 2\theta^{}_{12} \left(1 - 6 s^2_{13} + 9 s^4_{13}\right)
\nonumber \\
\hspace{-0.2cm} & & \hspace{-0.2cm}
+ \hspace{0.1cm} \frac{1}{8} \left[4\cos^2 2\theta^{}_{12}
\left(1 - 3 s^2_{13} + 3 s^4_{13}\right)
+ \sin^2 2\theta^{}_{12} c^4_{13}\right] \epsilon^2_\theta
+ \frac{1}{4} \sin^2 2\theta^{}_{12} \left(1 - 2 s^2_{13}
+ 3 s^4_{13}\right) \epsilon^2_\delta
\nonumber \\
\hspace{-0.2cm} & & \hspace{-0.2cm}
- \hspace{0.1cm} \frac{1}{4} \sin 4\theta^{}_{12} \left(1 - 3 s^2_{13} + 4 s^4_{13}\right)
\epsilon^{}_\theta \epsilon^{}_\delta \; .
\label{16}
\end{eqnarray}
These analytically exact results clearly show how sensitive $D^{}_0$ and
$C^{}_{\alpha\beta}$ are to the small $\mu$-$\tau$ symmetry breaking parameters
$\epsilon^{}_\theta$ and $\epsilon^{}_\delta$, consistent with our preliminary
expectations made below Eq.~(\ref{13}). Some further discussions are in order.
\begin{itemize}
\item     The ratio $C^{}_{ee}/D^{}_0$ is finite in most cases unless $D^{}_0$
itself vanishes, and its approximate result in the $\epsilon^{}_\theta \to 0$ or
$\epsilon^{}_\delta \to 0$ limit turns out to be
\begin{eqnarray}
\left. \frac{C^{}_{ee}}{D^{}_0}\right|_{\epsilon^{}_\theta \to 0} \simeq 3 \; ,
\quad
\left. \frac{C^{}_{ee}}{D^{}_0}\right|_{\epsilon^{}_\delta \to 0} \simeq
1 + \frac{1}{2} \tan^2 2\theta^{}_{12} \; , \hspace{1cm}
\label{17}
\end{eqnarray}
where the terms of ${\cal O}(s^2_{13})$ or smaller have been omitted. Of course,
the explicit value of $C^{}_{ee}/D^{}_0$ is highly sensitive to the inputs of
$\epsilon^{}_\theta$ and $\epsilon^{}_\delta$.

\item     In the leading-order approximation, the expressions of $C^{}_{e\mu}$
and $C^{}_{e\tau}$ can be simplified to
\begin{eqnarray}
C^{}_{e\mu} \simeq - C^{}_{e\tau} \simeq
-\frac{1}{4} \left(\sin^2 2\theta^{}_{12} \epsilon^{}_\theta + \sin^{} 4\theta^{}_{12}
\epsilon^{}_\delta\right) \; ,
\hspace{1cm}
\label{18}
\end{eqnarray}
where the terms of ${\cal O}(s^2_{13})$, ${\cal O}(\epsilon^{2}_\theta)$,
${\cal O}(\epsilon^{2}_\delta)$ or smaller have been neglected. Note, however,
that a significant cancellation between the two terms on the right-hand side of
Eq.~(\ref{18}) is likely to make the approximation unreliable, because
$\epsilon^{}_\theta$ and $\epsilon^{}_\delta$ may possess the opposite signs.

\item     In the leading-order approximation, the expressions of $C^{}_{\mu\mu}$,
$C^{}_{\tau\tau}$ and $C^{}_{\mu\tau}$ can be simplified to
\begin{eqnarray}
C^{}_{\mu\mu} \simeq C^{}_{\tau\tau} \simeq -C^{}_{\mu\tau} \simeq
\frac{1}{2} \cos^2 2\theta^{}_{12} + \frac{1}{8} \sin^2 2\theta^{}_{12} \; ,
\hspace{1cm}
\label{19}
\end{eqnarray}
where the terms of ${\cal O}(s^2_{13})$, ${\cal O}(\epsilon^{}_\theta)$,
${\cal O}(\epsilon^{}_\delta)$ or smaller have been omitted. So the
explicit values of $C^{}_{\mu\mu}/D^{}_0$, $C^{}_{\mu\tau}/D^{}_0$ and
$C^{}_{\tau\tau}/D^{}_0$ must be remarkably sensitive to the inputs of
$\epsilon^{}_\theta$ and $\epsilon^{}_\delta$, and their magnitudes must be
much larger than those of $C^{}_{ee}/D^{}_0$, $C^{}_{e\mu}/D^{}_0$ and
$C^{}_{e\tau}/D^{}_0$.
\end{itemize}
In any case, Eqs.~(\ref{15}) and (\ref{16}) provide us with a general and explicit
tool to infer the source flavor fractions $(\eta^{}_e, \eta^{}_\mu, \eta^{}_\tau)$
from the neutrino telescope data of $(f^{}_e , f^{}_\mu , f^{}_\tau)$ via Eq.~(\ref{11}).

For the sake of illustration, let us take the best-fit values
$s^2_{12} = 3.03 \times 10^{-1}$, $s^2_{13} = 2.23 \times 10^{-2}$,
$s^2_{23} = 4.73 \times 10^{-1}$ and $\delta^{}_\nu = 1.20 \times \pi$ as the
typical inputs~\cite{Capozzi:2025wyn} to calculate the coefficient matrices in
Eqs.~(\ref{1}) and (\ref{11}). In this case we obtain $\epsilon^{}_\theta \simeq 0.054$
and $\epsilon^{}_\delta \simeq -0.121$, and thus
\begin{eqnarray}
\left[\begin{matrix}
f^{}_e \cr f^{}_\mu \cr f^{}_\tau \end{matrix}\right]
\simeq \left(\begin{matrix}
0.5526 & \hspace{0.19cm} 0.2125 \hspace{0.19cm}  & 0.2348 \cr
0.2125 & 0.4078 & 0.3797 \cr
0.2348 & 0.3797 & 0.3855
\end{matrix}\right)
\left[\begin{matrix} \eta^{}_e \cr \eta^{}_\mu \cr
\eta^{}_\tau \end{matrix}\right] \; ,
\label{20}
\end{eqnarray}
and
\begin{eqnarray}
\left[\begin{matrix} \eta^{}_e \cr \eta^{}_\mu \cr
\eta^{}_\tau \end{matrix}\right]
\simeq \left(\begin{matrix}
2.505 & 1.392 & -2.897 \cr
1.392 & 30.41 & -30.80 \cr
-2.897 & -30.80 & 34.70
\end{matrix}\right)
\left[\begin{matrix} f^{}_e \cr f^{}_\mu \cr
f^{}_\tau \end{matrix}\right] \; ,
\label{21}
\end{eqnarray}
in which the $\mu$-$\tau$ symmetry breaking effects are appreciable.
Given $\eta^{}_e = 1/3$, $\eta^{}_\mu = 2/3$ and $\eta^{}_\tau = 0$
at a remote astrophysical source of high-energy neutrinos, for example,
we arrive at an approximate ``flavor democracy" at a
neutrino telescope with the help of Eq.~(\ref{20}): $f^{}_e \simeq 0.326$,
$f^{}_\mu \simeq 0.343$ and $f^{}_\tau \simeq 0.331$. In contrast,
making use of Eq.~(\ref{21}) to infer $\eta^{}_\mu$ and $\eta^{}_\tau$
from $f^{}_e$, $f^{}_\mu$ and $f^{}_\tau$ may involve significant
cancellations due to the very fact that the sizes of $C^{}_{\mu\mu}$
(or $C^{}_{\tau\tau}$) and $C^{}_{\mu\tau}$ are unusually large but their
signs are opposite --- a salient reflection of the approximate $\mu$-$\tau$
interchange symmetry hidden in the PMNS lepton flavor mixing matrix $U$
\footnote{Eq.~(\ref{5}) tells us that the exact $\mu$-$\tau$ symmetry
$\epsilon^{}_\theta = \epsilon^{}_\delta = 0$
leads to either $\theta^{}_{23} = \pi/4$ and $\theta^{}_{13} =0$ or
$\theta^{}_{23} = \pi/4$ and $\delta^{}_\nu = \pm \pi/2$. The former
has been ruled out by $\theta^{}_{13} \neq 0$~\cite{DayaBay:2012fng},
but the latter remains viable (especially for $\delta^{}_\nu =
-\pi/2$~\cite{T2K:2025wet}). Because the most poorly known
CP-violating phase $\delta^{}_\nu$ is always associated with
$s^{}_{13}$ in the standard parametrization of the PMNS matrix $U$, it is
actually $\theta^{}_{23} \sim \pi/4$ and $s^{}_{13} \ll 1$ that assure
the approximate $\mu$-$\tau$ flavor symmetry relations
$|U^{}_{\mu i}| \simeq |U^{}_{\tau i}|$ (for $i = 1, 2, 3$) to hold.}.

To give one a ball-park feeling of the potential divergence in inferring
the source flavor fractions $\eta^{}_\mu$ and $\eta^{}_\tau$ from the
measurements of $(f^{}_e , f^{}_\mu , f^{}_\tau)$ at
a neutrino telescope, let us apply Eq.~(\ref{21}) to the recent
IceCube all-sky neutrino flux data ranging from 5 TeV to 10 PeV by simply
assuming that all the relevant neutrino sources have a common flavor
composition. In this case the IceCube best-fit flavor ratios $f^{}_e = 0.30$,
$f^{}_\mu = 0.37$ and $f^{}_\tau = 0.33$~\cite{Abbasi:2025fjc} result in
the following flavor composition at those sources: $\eta^{}_e \simeq 0.31$,
$\eta^{}_\mu \simeq 1.50$ and $\eta^{}_\tau \simeq -0.81$. The individual
outputs of $\eta^{}_\mu$ and $\eta^{}_\tau$ are certainly meaningless, but
their sum $\eta^{}_\mu + \eta^{}_\tau \simeq 0.69 \simeq 2.2 \eta^{}_e$ makes
sense.

The above problem is likely to originate from our naive assumption of the
same flavor composition for all the astrophysical neutrino sources under
discussion, but it is more likely to arise from the uncertainties associated
with the present IceCube data about $f^{}_\mu$ and $f^{}_\tau$. Such
uncertainties may be significantly amplified by the approximate $\mu$-$\tau$
flavor symmetry of $U$ and hence obstruct a meaningful numerical extraction
of $\eta^{}_\mu$ and $\eta^{}_\tau$
\footnote{In this connection, a careful numerical analysis as done in
Ref.~\cite{Bustamante:2019sdb} is necessary to examine how the inferred
source flavor composition is sensitive to the uncertainties associated
with both the neutrino oscillation parameters and the Earth flavor composition
observed at a neutrino telescope within the standard three-flavor neutrino
mixing scheme, and to discern whether a seemingly meaningless output might
actually be an indication of possible new physics. We are going to carry
out a comprehensive numerical study of this kind by using the latest
experimental data elsewhere.}.
That is why a sufficiently accurate measurement of the flavor distributions
$f^{}_\alpha$ (for $\alpha = e, \mu, \tau$) at the neutrino telescopes is
highly desired, in order to successfully determine the source flavor fractions
$\eta^{}_\alpha$ (for $\alpha = e, \mu, \tau$)~\cite{IceCube-Gen2:2020qha,
KM3NeT:2018wnd,TRIDENT:2022hql,Chen:2026jdx}.

\section{In the $\mu$-$\tau$ symmetry limit}

Now we come back to the limit of an exact $\mu$-$\tau$ flavor symmetry
associated with the PMNS lepton flavor mixing matrix $U$, as this
special case remains consistent with current neutrino oscillation
data at least at the $2\sigma$ confidence level~\cite{Capozzi:2025wyn,
Capozzi:2025ovi,Esteban:2024eli,Esteban:2026phq}. Taking
$\epsilon^{}_\theta = \epsilon^{}_\delta = 0$, which is equivalent to
$\theta^{}_{23} = \pi/4$ and $\delta^{}_\nu = -\pi/2$ in the standard
parametrization of $U$, we arrive at $P^{}_{e\mu} = P^{}_{e\tau}$ and
$P^{}_{\mu\mu} = P^{}_{\mu\tau} = P^{}_{\tau\tau}$
from Eq.~(\ref{2}). As a consequence, the three linear equations in
Eq.~(\ref{6}) are reduced to two independent linear equations in
two variables $\eta^{}_e$ and $\left(\eta^{}_\mu + \eta^{}_\tau\right)$:
\begin{eqnarray}
P^{}_{ee} \eta^{}_e + P^{}_{e\mu} \left(\eta^{}_\mu + \eta^{}_\tau\right)
\hspace{-0.2cm} & = & \hspace{-0.2cm} f^{}_e \; ,
\nonumber \\
P^{}_{e\mu} \eta^{}_e + P^{}_{\mu\mu} \left(\eta^{}_\mu + \eta^{}_\tau\right)
\hspace{-0.2cm} & = & \hspace{-0.2cm} f^{}_\mu \; , \hspace{0.5cm}
\label{22}
\end{eqnarray}
as $f^{}_\tau = f^{}_\mu$ holds in this symmetry limit~\cite{Fu:2012zr,Fu:2014isa}.
Then we obtain
\begin{eqnarray}
&& \eta^{}_\mu + \eta^{}_\tau =
\frac{P^{}_{ee} f^{}_\mu - P^{}_{e\mu} f^{}_e}{P^{}_{ee} P^{}_{\mu\mu} - P^2_{e\mu}}
= \frac{4\left(2 - \sin^2 2\theta^{}_{12} c^4_{13} - \sin^2 2\theta^{}_{13}
\right) f^{}_\mu - 2 \left(\sin^2 2\theta^{}_{12} c^4_{13} + \sin^2 2\theta^{}_{13}
\right) f^{}_e}{4 - 3 \sin^2 2\theta^{}_{12} c^4_{13} - 3 \sin^2 2\theta^{}_{13}}
\; , \hspace{0.5cm}
\nonumber \\
&& \eta^{}_e =
\frac{P^{}_{\mu\mu} f^{}_e - P^{}_{e\mu} f^{}_\mu}
{P^{}_{ee} P^{}_{\mu\mu} - P^2_{e\mu}}
= \frac{\left(4 - \sin^2 2\theta^{}_{12} c^4_{13} - \sin^2 2\theta^{}_{13}\right)
f^{}_e - 2 \left(\sin^2 2\theta^{}_{12} c^4_{13} + \sin^2 2\theta^{}_{13}\right)
f^{}_\mu}{4 - 3 \sin^2 2\theta^{}_{12} c^4_{13} - 3 \sin^2 2\theta^{}_{13}} \; ,
\label{23}
\end{eqnarray}
where $\epsilon^{}_\theta = \epsilon^{}_\delta = 0$ have
been input. It is clear that $\eta^{}_e + \eta^{}_\mu + \eta^{}_\tau
= f^{}_e + 2 f^{}_\mu$ holds, and the exact $\mu$-$\tau$ flavor symmetry
only allows us to infer a sum of the $\mu$- and $\tau$-flavor fractions
of high-energy neutrinos at an astrophysical source from
the observed flavor ratios at a neutrino telescope.

It is certainly improper to apply the global-analysis best-fit
values of $\theta^{}_{12}$ and $\theta^{}_{13}$ to Eq.~(\ref{23}),
as they are inconsistent with the exact
$\mu$-$\tau$ symmetry conditions. But direct measurements of
$\theta^{}_{12}$ and $\theta^{}_{13}$ in the JUNO and Daya Bay
experiments have nothing to do with $\theta^{}_{23}$ and
$\delta^{}_\nu$, and thus they are applicable to the numerical
calculations of $\eta^{}_e$ and $\eta^{}_\mu + \eta^{}_\tau$
by means of Eq.~(\ref{23}). Given the latest
best-fit results $s^2_{12} = 3.092 \times 10^{-1}$ and
$s^2_{13} = 2.175 \times 10^{-2}$ reported respectively by the
JUNO~\cite{JUNO:2025gmd} and Daya Bay~\cite{DayaBay:2022orm}
Collaborations, we immediately arrive at
\begin{eqnarray}
\eta^{}_e \simeq 2.398 f^{}_e - 1.398 f^{}_\mu \; , \quad
\eta^{}_\mu + \eta^{}_\tau \simeq 3.398 f^{}_\mu - 1.398 f^{}_e \; .
\label{24}
\end{eqnarray}
So the ``flavor democracy" $f^{}_e = f^{}_\mu = f^{}_\tau =1/3$
at a neutrino telescope will automatically lead us to
$\eta^{}_e = 1/3$ and $\eta^{}_\mu + \eta^{}_\tau = 2/3$ at the
astrophysical source.

Note, however, that the applicability of the above discussions
are subject to two prerequisites: first, the observed
flavor ratios $f^{}_\mu$ and $f^{}_\tau$ should satisfy $f^{}_\mu = f^{}_\tau$
to a good degree of accuracy; second, the $\mu$-$\tau$ flavor
symmetry should be allowed by the precision neutrino oscillation
data. As both of them are rather restrictive, we
expect that a realistic telescope-to-source approach must involve
small $\mu$-$\tau$ symmetry breaking effects as well as certain divergence
and ambiguities in extracting the original flavor composition of
high-energy astrophysical neutrinos.

\section{Summary}

With no assumption of any specific flavor composition pattern
for a remote astrophysical source of high-energy neutrinos,
we have calculated their original flavor fractions
$(\eta^{}_e , \eta^{}_\mu , \eta^{}_\tau)$ in terms of the flavor ratios
$(f^{}_e , f^{}_\mu , f^{}_\tau)$ observed at a neutrino telescope
within the standard three-flavor oscillation scheme. In particular,
we have presented a complete set of analytical expressions for
$(\eta^{}_e , \eta^{}_\mu , \eta^{}_\tau)$ as functions of two typical $\mu$-$\tau$
symmetry breaking parameters in the standard parametrization of the PMNS
matrix $U$. We find that a potential divergence is expected to appear in inferring
the values of $\eta^{}_\mu$ and $\eta^{}_\tau$, as an unavoidable
consequence of the observed approximate $\mu$-$\tau$ interchange
symmetry of $U$. We have illustrated
this effect by applying our formulas to the recent IceCube all-sky neutrino
flux data ranging from 5 TeV to 10 PeV by assuming that the
relevant sources have a common flavor composition. In addition, we
have explained why only $\eta^{}_e$ and $\eta^{}_\mu + \eta^{}_\tau$
can be extracted from the precision measurements of $f^{}_e$ and
$f^{}_\mu = f^{}_\tau$ if the $\mu$-$\tau$ flavor symmetry is exact.

We conclude that a precise determination of the $\mu$-$\tau$ flavor
symmetry breaking effects in the upcoming long-baseline accelerator
neutrino oscillation experiments is crucial, in order to more reliably
trace the true flavor distribution of high-energy astrophysical
neutrinos with the precision measurements at a number of neutrino telescopes.

\section*{Acknowledgements}

This research was supported in part by the National Natural Science Foundation
of China under Grant No. 12535007, and by the Scientific and Technological
Innovation Program of the Institute of High Energy Physics under Grant
No. E55457U2.


\end{document}